\documentclass[twocolumn]{aastex7} 
\usepackage{amsmath}
\usepackage{float}
\usepackage{graphicx}
\usepackage{amssymb}
\usepackage{amsbsy}
\usepackage{amsmath}
\usepackage{amsfonts}
\usepackage{mathrsfs}
\usepackage{color}
\usepackage{multirow}
\usepackage{booktabs}	
\usepackage{diagbox}
\usepackage{booktabs}              
\usepackage{makecell}      
\usepackage{array}        
\usepackage{rotating} 

\usepackage{bm}
\DeclareMathAlphabet{\mathbi}{OT1}{ptm}{bx}{it}
\SetMathAlphabet\mathbi{bold}{OT1}{ptm}{bx}{it}

\shorttitle{SMBH and BLR in NGC~5548}
\shortauthors{Xi et al.}

\begin{document}

\title{\bf\large Supermassive Black Hole and Broad-line Region in NGC~5548: 2023 Reverberation Mapping Results}

\author{Wen-Zhe Xi}
\affiliation{Yunnan Observatories, Chinese Academy of Sciences, Kunming 650011, People's Republic of China} 
\affiliation{University of Chinese Academy of Sciences, Beijing 100049, People's Republic of China}
\email{xiwenzhe@ynao.ac.cn}

\author[0000-0002-2310-0982]{Kai-Xing Lu}
\affiliation{Yunnan Observatories, Chinese Academy of Sciences, Kunming 650011, People's Republic of China} 
\affiliation{University of Chinese Academy of Sciences, Beijing 100049, People's Republic of China}
\affiliation{Key Laboratory for the Structure and Evolution of Celestial Objects, Chinese Academy of Sciences, Kunming 650011, People's Republic of China} 
\email[show]{lukx@ynao.ac.cn}

\author{Jin-Ming Bai}
\affiliation{Yunnan Observatories, Chinese Academy of Sciences, Kunming 650011, People's Republic of China}
\affiliation{University of Chinese Academy of Sciences, Beijing 100049, People's Republic of China}
\affiliation{Key Laboratory for the Structure and Evolution of Celestial Objects, Chinese Academy of Sciences, Kunming 650011, People's Republic of China} 
\email[show]{baijinming@ynao.ac.cn}

\author{Zhang Yue}
\affiliation{Yunnan Observatories, Chinese Academy of Sciences, Kunming 650011, People's Republic of China} 
\affiliation{University of Chinese Academy of Sciences, Beijing 100049, People's Republic of China}
\email{1398611241@qq.com}

\author{Weimin Yi}
\affiliation{Yunnan Observatories, Chinese Academy of Sciences, Kunming 650011, People's Republic of China} 
\affiliation{Key Laboratory for the Structure and Evolution of Celestial Objects, Chinese Academy of Sciences, Kunming 650011, People's Republic of China} 
\email{}

\author{Liang Xu}
\affiliation{Yunnan Observatories, Chinese Academy of Sciences, Kunming 650011, People's Republic of China}
\affiliation{Key Laboratory for the Structure and Evolution of Celestial Objects, Chinese Academy of Sciences, Kunming 650011, People's Republic of China} 
\email{xuliang@ynao.ac.cn}

\author[0000-0003-3823-3419]{Sha-Sha Li}
\affiliation{Yunnan Observatories, Chinese Academy of Sciences, Kunming 650011, People's Republic of China} 
\affiliation{Key Laboratory for the Structure and Evolution of Celestial Objects, Chinese Academy of Sciences, Kunming 650011, People's Republic of China} 
\email{lishasha@ynao.ac.cn}

\author[0000-0002-1530-2680]{Hai-Cheng Feng}
\affiliation{Yunnan Observatories, Chinese Academy of Sciences, Kunming 650011, People's Republic of China} 
\affiliation{Key Laboratory for the Structure and Evolution of Celestial Objects, Chinese Academy of Sciences, Kunming 650011, People's Republic of China} 
\email{hcfeng@ynao.ac.cn}

\author[0000-0003-4156-3793]{Jian-Guo Wang}
\affiliation{Yunnan Observatories, Chinese Academy of Sciences, Kunming 650011, People's Republic of China} 
\affiliation{Key Laboratory for the Structure and Evolution of Celestial Objects, Chinese Academy of Sciences, Kunming 650011, People's Republic of China} 
\email{wangjg@ynao.ac.cn}

\begin{abstract}
We present the results of the 2023 spectroscopic reverberation mapping (RM) campaign for active galactic nuclei (AGN) of NGC 5548, 
continuing our long-term monitoring program. 
Using the Lijiang 2.4-meter telescope, we obtained 74 spectra with a median cadence of 1.9 days. 
Through detailed spectral decomposition, we measured the light curves of the optical continuum at 5100~\AA\ and 
the broad He~{\sc ii}, He~{\sc i}, H$\gamma$, and H$\beta$ emission lines. 
The time lags of these lines relative to the continuum are measured as $1.3^{+1.6}_{-0.6}$, $2.3^{+1.5}_{-2.1}$, 
$10.0^{+2.0}_{-1.8}$, and $15.6^{+2.6}_{-2.9}$ days (rest-frame), respectively. 
Velocity-resolved lag profiles for H$\gamma$ and H$\beta$ were constructed. 
Combined with data from previous seasons (2015$-$2021), 
we find that the radial ionization stratification of the broad-line region (BLR) is stable; 
the average virial mass of the supermassive black hole in NGC~5548 is $(2.6\pm1.1)\times 10^{8}M_{\odot}$, 
consistent with the $M_{\rm BH}-\sigma_*$ relation; 
the broad He~{\sc ii} line exhibits the largest responsivity, followed by broad He~{\sc i} (or H$\gamma$) and H$\beta$ lines; 
the BLR kinematics show significant temporal evolution, transitioning from virialized motions to signatures of inflow and outflow. 
Furthermore, an analysis of 35 years of historical data confirms a 3.5-year time lag between variations 
in the optical luminosity and the BLR radius, potentially implicating the role of radiation pressure 
or dynamical structure changes in the inner accretion disk. 
Long-term campaign demonstrates that the BLR in NGC 5548 is a robust yet dynamically evolving entity, 
providing crucial insights into AGN structure and accretion physics. 
\end{abstract}


\section{Introduction} \label{sec:intro}

Reverberation mapping (RM; \citealt{Blandford1982,Peterson1993}) is a powerful tool to probe the kinematics of the broad-line region (BLR) 
and measure masses of the accreting supermassive black hole (SMBH) in the centres of active galactic nuclei 
(AGN; e.g., \citealt{Peterson1999,Bentz2009,Denney2010,Du2018,Lu2021a,Hu2021,Rakshit2019,Woo2019,Woo2024,Cho2023,Wang2025}). 
Since 1989 \citep{Peterson1993}, 
NGC~5548 (14:17:59.534, +25:08:12.44, $z=0.01717$) as a nearby Seyfert galaxy, has been monitored with long-term spectroscopic RM campaigns, 
to understand its BLR properties and evolution (\citealt{Peterson2002,Bentz2009,Denney2010,Pei2017,DeRosa2018,Lu2016,Lu2022}), 
and further to understand BLR physical properties across the AGN population. 
\cite{Lu2016,Lu2022} found that the BLR radius responds to changes in optical luminosity with a time delay of 3.5~years. 
They suggested that such a time delay might be related to the radiation pressure from the central accretion disc. 
Meanwhile, \cite{Li2016} and \cite{Bon2016} detected periodic variations of the continuum and double-peaked profiles 
of the broad H$\beta$ line in NGC~5548, perhaps indicative of a sub-parsec SMBH binary candidate. 
Besides, a rare phenomenon called ``BLR holiday'', namely, the broad emission lines in variability decorrelate from the AGN continuum, 
was observed in 2014 (\citealt{Goad2016,Pei2017}). 
The falling corona model (\citealt{Sun2018}) and disc wind model (\citealt{Dehghanian2020}) 
were proposed to explain this anomalous behavior, yet without having reached a consensus. 
Given that NGC~5548 has undergone extreme variability in the past two decades (e.g., \citealt{Li2016}), 
it is a potential candidate for a changing-look AGN. 
Long-term and continuous RM observations for the BLR of NGC 5548 
may offer crucial diagnostics to pin down the changing-look behaviors in the future. 
Taken together, investigating the kinematics of the BLR in NGC 5548 and its variations over a long timescale is highly worthwhile. 

In addition, the geometry and dynamics of the BLR in NGC 5548 have been studied via velocity-resolved reverberation mapping, 
including velocity-binned RM (\citealt{Bentz2009,Lu2016,Lu2022}), dynamical modeling (\citealt{Williams2020}), 
and MEMEcho analysis (\citealt{Horne1994,Xiao2018,Horne2021}). 
\cite{Horne2021} used AGN STORM data to construct the first high-resolution velocity-delay maps 
for Ly$\alpha$, C~{\sc iv}, He~{\sc ii}, and H$\beta$, revealing the BLR’s structure, kinematics, and ionization stratification. 
Notably, MEMEcho residuals revealed a helical ``Barber-Pole" pattern shifting from red to blue 
across C~{\sc iv} and Ly$\alpha$ with a $\sim$2-year period, 
suggesting an azimuthal structure rotating on the far side of the accretion disk, 
possibly due to precession or orbital motion. 
Long-term monitoring of this pattern could offer a promising way to probe inner disk dynamics. 

Therefore, NGC~5548 was the highest priority target in our long-term spectroscopic monitoring campaign (\citealt{Lu2022}). 
In this work, we report the analysis and results of RM campaign from the 2023 observing season. 
The paper is organized as follows. Section~\ref{sec:od} describes the observations and data reduction. 
Section~\ref{sec:data} presents data analysis, including the measurements of light curves, time lags, and line widths. 
In Section~\ref{sec:mass}, we estimate the viral mass of supermassive black in NGC 5548. 
Section~\ref{sec:blr} presents the BLR properties, including ionization stratification and kinematics of BLR. 
We end with a summary of our main results in Section~\ref{sec:sum}. 

\section{Observation and Data Reduction} \label{sec:od}
After the work of \cite{Lu2022}, we conducted additional spectroscopic monitoring of NGC~5548 in 2023.
The spectroscopic observation settings and data reduction procedures were identical to those described in \citep{Lu2016,Lu2022}. 
For more detailed discussions on RM experiments, 
readers are referred to our previous studies on other AGNs \citep{Lu2019a,Lu2021a}. 

\subsection{Spectroscopy} \label{sec:so}
Our spectroscopic observations of NGC~5548 were taken using 
Yunnan Faint Object Spectrograph and Camera (YFOSC) mounted on the Lijiang 2.4~m telescope (LJT), 
which locates in the Lijiang observatory of Yunnan Observatories, Chinese Academy of Sciences. 
YFOSC is equipped with a  back-illuminated 2048$\times$2048 pixel CCD, 
with the pixel size 13.5 $\mu$m, pixel scale 0.283$''$ per pixel, and field-of-view $10'\times10'$. 
More information about the Lijiang observatory and telescope is described in \cite{Fan2015}, \cite{Wang2019}, \cite{Xin2020} and \cite{Lu2021b}. 

Following the previous observations in \cite{Lu2022}, 
we oriented a long slit to take the spectra of NGC~5548 and a previously selected non-varying comparison star 
(SDSS~J141758.82+250533.1, hereafter J1417) simultaneously. 
J1417 is a G1-type star with a $V$-band magnitude of 13.9. 
Our previous $g$-band light curve confirms J1417's stability (see Figure 1 of \citealt{Lu2022}), making it an ideal reference star for NGC~5548. 
This observation strategy,  detailed in \citet{Maoz1990} and \citet{Kaspi2000}, has been adopted in some RM campaigns (e.g., \citealt{Du2015,Lu2021a}). 
Our prior works show that the comparison star's spectrum can precisely calibrate spectral flux (\citealt{Lu2019a}, see also \citealt{Hu2015}), 
and correct telluric absorption lines of the target's spectra (\citealt{Lu2021b}). 
Based on the average seeing at Lijiang Observatory, 
we adopted a long slit with width of $2.5''$. We used Grism 14 which covers the wavelength range $\sim$3600~\AA\ to 7460~\AA\ and provides a dispersion of 1.8~\AA\,~pixel$^{-1}$. 
The standard neon and helium lamps were used for wavelength calibration. 

\subsection{Data Reduction}\label{sec:dr}
We obtained 74 spectroscopic observations during the 2023 observing season, 
spanning from January to June with a median sampling interval of 1.9 days. 
The average S/N ratio of the spectra at 5100~\AA\ is 73 per 10-pixel resolution element, with a standard deviation of 22. 
Two-dimensional spectra were reduced using the standard {\tt IRAF} procedures, 
including bias subtraction, flat-field correction, wavelength calibration, and cosmic ray elimination. 
A relatively small extraction window of 20 pixels (5.7$^{\prime\prime}$) was used to reduce  Poisson noise 
from the sky background and enhance the signal-to-noise (S/N) ratio. 
Sky background was subtracted based on two adjacent regions 
$\pm$($7.4^{\prime\prime}$ to $14^{\prime\prime}$) outside the extraction window.

NGC~5548 and its comparison star were observed simultaneously under the same conditions (e.g., airmass and seeing). 
The spectrum of NGC~5548 was calibrated using the sensitivity function derived from the comparison star (see \citealt{Lu2019a}). 
Following previous studies (\citealt{Du2015,Lu2016,Lu2021a}), we generated the fiducial spectrum of the comparison star from photometric nights and calculated the sensitivity function by comparing it to the observed spectra. This function was then applied to calibrate NGC~5548's spectrum. 
Galactic extinction was corrected using \cite{Schlegel1998}'s extinction map, 
and wavelength shifts due to varying seeing or mis-centering were corrected using the [O~{\sc iii}]~$\lambda5007$ line as a reference. 
Finally, all spectra were transformed into the rest frame for analysis. 

\begin{figure*}[htb]
\centering
\includegraphics[angle=0,width=0.99\textwidth]{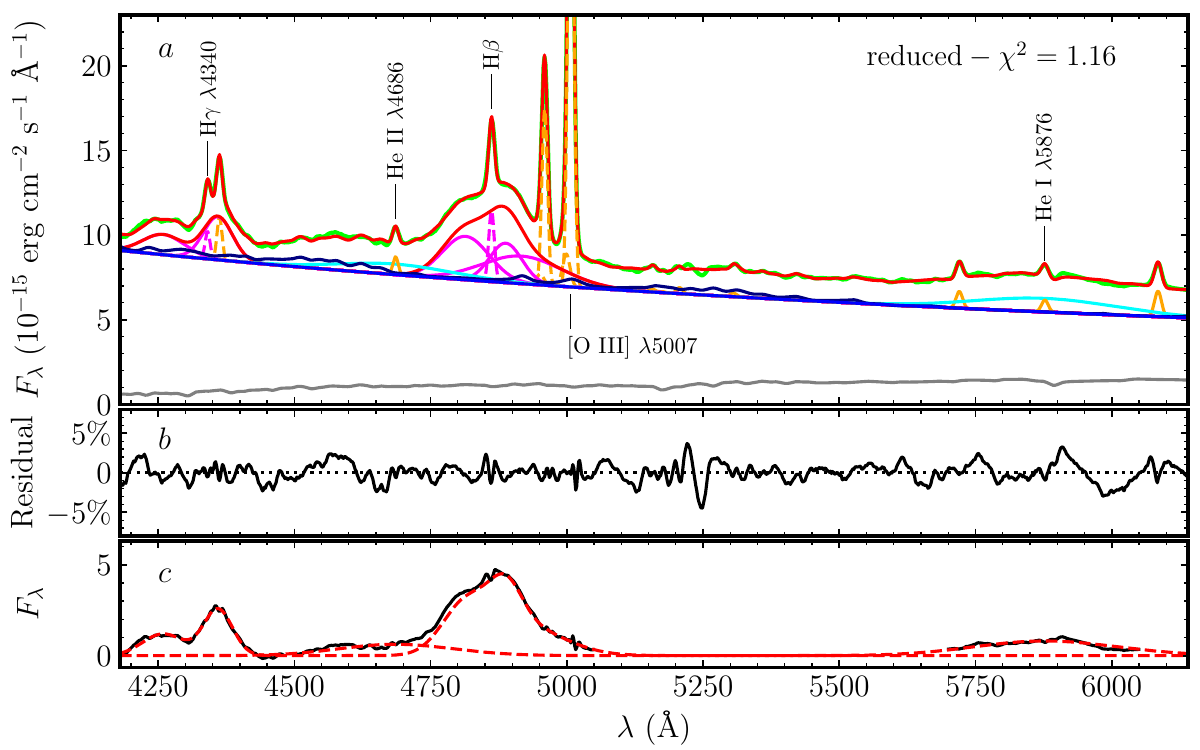}
\caption{\footnotesize
Spectral fitting and decomposition of the mean spectrum calculated from the calibrated spectra. 
Panel~({\it a}) illustrates the details of spectral fitting and decomposition, where the mean spectrum is shown in lime, the total model in red, 
and the fitting components include the AGN continuum (blue), iron multiplets (navy), host galaxy (gray), broad Balmer lines (solid magenta), 
narrow Balmer lines (dashed magenta), broad helium lines (cyan), narrow helium lines (orange), [O~{\sc iii}] narrow lines (dashed orange), 
and other weak narrow emission lines (solid orange). 
Vertical reference lines are added at the rest wavelengths 
to mark the broad H$\gamma$, He~{\sc ii}, H$\beta$, He~{\sc ii}, and [O~{\sc iii}]~$\lambda5007$ lines. 
Panel~({\it b}) displays the fitting residuals in percentage. 
Panel~({\it c}) presents the net broad H$\gamma$, He~{\sc ii}, H$\beta$, and He~{\sc i} lines, 
where the fitted broad-line profiles (models) are indicated by red dashed lines, 
and the broad-line profiles after subtracting other fitted components from the spectrum are shown in black. 
}
\label{fig_fit1}
\end{figure*}

\begin{deluxetable*}{cccccc}
\tablecolumns{6}
\tabletypesize{\scriptsize}
\setlength{\tabcolsep}{18pt}
\tablewidth{4pt}
\tablecaption{Light Curves\label{tab_lc}}
\tablehead{
\colhead{JD}                     &
\colhead{$F_{\rm 5100}$}          &
\colhead{$F_{\rm He~II}$}         &
\colhead{$F_{\rm He~I}$}          &
\colhead{$F_{\rm H\gamma}$}  &
\colhead{$F_{\rm H\beta}$}
}
\startdata
2459925.4 &$6.432 \pm0.128 $&$1.347 \pm0.139 $&$2.708 \pm0.125 $&$2.595 \pm0.105 $&$7.379 \pm0.143$ \\
2459937.4 &$6.237 \pm0.127 $&$1.209 \pm0.137 $&$2.668 \pm0.124 $&$2.849 \pm0.103 $&$7.606 \pm0.144$ \\
2459939.4 &$6.253 \pm0.129 $&$1.594 \pm0.143 $&$2.488 \pm0.126 $&$2.822 \pm0.107 $&$7.589 \pm0.144$ 
\enddata
\tablecomments{\footnotesize
The 5100~\AA\ continuum flux is given in units of ${\rm 10^{-15}~erg~s^{-1}~cm^{-2}~\AA^{-1}}$,  
and the broad H$\gamma$, He~{\sc ii}, H$\beta$, and He~{\sc i} lines are presented in units of ${\rm 10^{-13}~erg~s^{-1}~cm^{-2}}$.  
The contamination from the host galaxy has been eliminated. 
(This table is available in its entirety in machine-readable form.)
}
\end{deluxetable*}

\begin{deluxetable*}{cccccc}
\tablecolumns{6}
\tabletypesize{\scriptsize}
\setlength{\tabcolsep}{18pt}
\tablewidth{4pt}
\tablecaption{Light-curve Statistics \label{tab_lcstat}}
\tablehead{
\colhead{Parameters}                     &
\colhead{$F_{\rm 5100}$}          &
\colhead{$F_{\rm He~II}$}         &
\colhead{$F_{\rm He~I}$}          &
\colhead{$F_{\rm H\gamma}$}  &
\colhead{$F_{\rm H\beta}$}
}
\startdata
Mean flux              &$6.69\pm0.52   $&$1.63\pm0.59   $&$2.65\pm0.28 $&$2.85\pm0.33   $&$7.53\pm0.47$ \\
$F_{\rm var}$(\%) &$7.43\pm 0.67 $&$35.39\pm 3.11$&$9.37\pm 1.00$&$10.69\pm 0.99$&$5.92\pm 0.54$ 
\enddata
\tablecomments{\footnotesize
The units of mean flux are the same as those in Table~\ref{tab_lc}.  
For each light curve, we calculated the mean flux and its uncertainty (the standard deviation of the light curve),  
as well as the variability amplitude $F_{\rm var}$ and its uncertainty (see Section~\ref{sec:lc}). 
}
\end{deluxetable*}

\section{Data Analysis}\label{sec:data}
\subsection{Light Curves}\label{sec:lc}
To disentangle blended components in AGN spectra, spectral fitting and decomposition are commonly used (\citealt{Hu2008,Bian2010,Dong2011,Barth2013,Guo2014,Barth2015,Lu2019b,Lu2025,Shen2025}). 
In particular, during RM campaigns of local AGNs, host-galaxy starlight dilutes AGN continuum and broad emission line variability and introduces random noise due to nightly seeing variations and slit misalignment (\citealt{Hu2015,Lu2019a}). 
We therefore measure the fluxes of the AGN continuum at 5100~\AA\ and broad He~{\sc ii}, He~{\sc i}, H$\gamma$, and H$\beta$ lines 
using spectral fitting and decomposition. 

Following previous works (\citealt{Hu2015,Lu2021a}), we fit and decompose calibrated spectra (Section~\ref{sec:dr}) 
using the MPFIT package (\citealt{Markwardt2009}), which minimizes $\chi^{2}$ via the Levenberg-Marquardt technique. 
The fitting window spans 4200$-$6110~\AA~(rest-frame), excluding wavelengths $>$6200~\AA\ due to second-order spectrum contamination. 
Components included in the fitting are:  
(1) A power law ($f_{\lambda}\propto\lambda^{\alpha}$, where $\alpha$ is the spectral index) for the AGN continuum.  
(2) An iron template from \cite{Boroson1992} for iron multiplets. Despite multiple templates available (e.g., \citealt{Boroson1992,Veron2004,Kova2010,Park2022}), NGC~5548's weak iron multiplets make our results insensitive to the specific choice.  
(3) Host-galaxy starlight modeled with a stellar template (age$=11$~Gyr, metallicity $Z=0.05$, \citealt{Bruzual2003}). 
Stellar absorption lines near 5876~\AA\ and 5180~\AA\ constrain the host-galaxy contribution.  
(4) Three Gaussians for broad H$\beta$ and H$\gamma$ lines.  
(5) Two double Gaussians for [O~{\sc iii}] $\lambda5007/\lambda4959$. 
(6) A Gaussian for narrow Balmer lines. 
(7) Single Gaussians for other narrow emission lines. 
All components are fitted simultaneously. 

We calculate the mean spectrum (\citealt{Peterson2004}) from above processed spectra and fit the mean spectrum before fitting the individual spectra. 
During mean spectrum fitting: 
the [O~{\sc iii}] doublet flux ratio is fixed at 3; 
the [O~{\sc iii}] $\lambda5007/\lambda4959$ and all other narrow lines share the same velocity and shift; 
different host-galaxy templates from \cite{Bruzual2003} are considered, 
however, the template with 11 Gyr age and metallicity $Z=0.05$ gives 
a plausible fit to the stellar absorption lines and the spectral index of 
AGN continuum ($\sim\lambda^{-1.5}$); remaining parameters vary freely. 
Figure~\ref{fig_fit1} displays the results of spectral fitting 
and decomposition of the mean spectrum. 
The panel ({\it a}) shows the details of spectral fitting and decomposition, along with the reduced-$\chi^{2}$. 
The panel ({\it b}) shows the fitting residuals in percentage. 
The panel ({\it c}) shows the net broad H$\gamma$, He~{\sc ii}, H$\beta$, and He~{\sc i} lines, 
where the best-fitted broad lines are in red dashed line and the total broad-line profiles obtained by subtracting other fitted components are in black. 
From this we can inspect the general shapes of the broad-line profiles. 
For individual spectrum fitting:  
(1) The spectral index ($\alpha$) is fixed to the value from the mean spectrum to reduce degeneracy between the power-law continuum and host galaxy. 
To check the impact of $\alpha$-fixed and $\alpha$-freed fitting on the results, 
we take a comparison between the $\alpha$-fixed and $\alpha$-freed light curves (including cross-correlation analysis) and find that both results are consistent. 
(2) Narrow lines are also tied with the same velocity and shift; 
(3) Broad He~{\sc ii} line width is fixed to the best-fit value from the mean spectrum due to its weak strength. 

\begin{figure*}[htb]
\centering
\includegraphics[angle=0,width=0.95\textwidth]{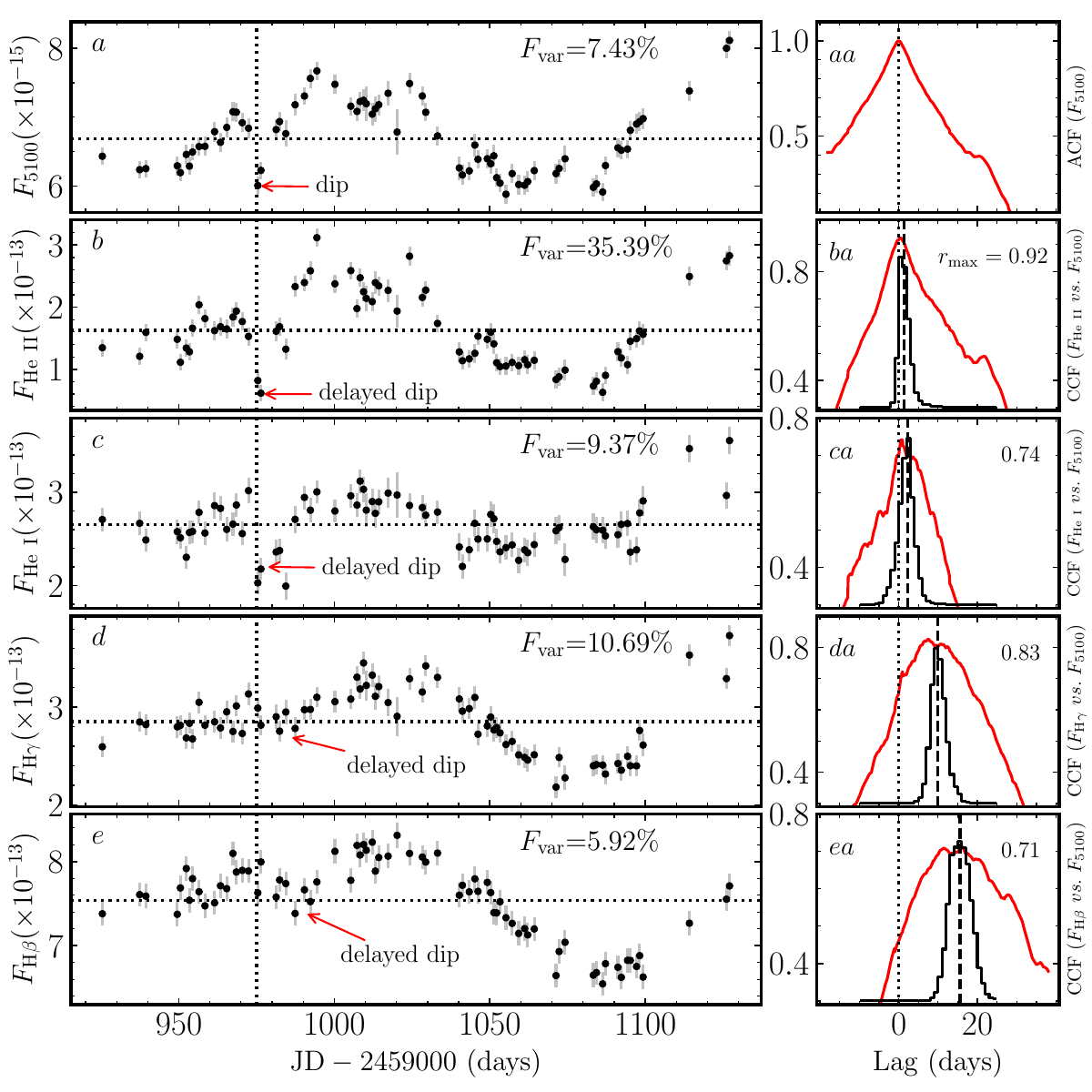}
\caption{\footnotesize
Light curves and the results of cross-correlation analysis. 
The left panels ({\it a-e}) are the light curves of the AGN continuum at 5100~\AA~and the broad He~{\sc ii}, He~{\sc i}, 
H$\gamma$, and H$\beta$ emission lines, the contamination of this data by the host galaxy has been eliminated by spectral decomposition. 
The red arrows indicate a transient dip in the AGN continuum and the corresponding, 
time-delayed dip in the emission-line flux. 
The horizontal dotted lines represent the mean flux. 
The right panels ({\it aa-ea}) are corresponding to the ACF of the continuum and the CCF between the broad-line light curves ({\it b-e}) and the continuum variation ({\it a}), the histogram in black is the cross-correlation centroid distribution (CCCD). 
We note the variability amplitude of $F_{\rm var}$ in panels~({\it a-e}), 
and the maximum cross-correlation coefficient of $r_{\max}$ in CCF panels~({\it ba-ea}). 
The measured time lags in rest-frame are marked by the vertical dashed lines in panels~({\it ba-ea}), 
the vertical dotted lines are reference lines of zero time lag. 
The units of $F_{\rm 5100}$ and emission lines are ${\rm erg~s^{-1}~cm^{-2}~\AA^{-1}}$ and ${\rm erg~s^{-1}~cm^{-2}}$, respectively. 
}
\label{fig_map2023}
\end{figure*}

For each fitting, we calculate the reduced-$\chi^{2}$, with a median value of 1.63 across all fits. 
Fluxes of the AGN continuum ($F_{\rm 5100}$) and broad He~{\sc ii}, He~{\sc i}, H$\gamma$, and H$\beta$ lines are measured from the best-fitted models. 
Light curves and uncertainties (including Poisson and systematic errors) are summarized in Table~\ref{tab_lc} and illustrated in Figure~\ref{fig_map2023}. 
A visual inspection of the overall light curve reveals a transient dip in the AGN continuum 
around the observation epoch corresponding to JD$-$2459000$=$975, which is marked by red arrows. 
After accounting for the time delay (see Section~\ref{sec:lag}), a corresponding, 
delayed dip is detected in the emission-line light curve, 
where the variation associated with the Balmer emission line exhibits a smoothed response. 

For each light curve, we compute the mean flux and variability amplitude ($F_{\rm var}$) as defined by \citet{Rodriguez-Pascual1997}:  
\begin{equation}
F_{\rm var}=\frac{\left(\sigma^2-\Delta^2\right)^{1/2}}{\langle F\rangle}, 
\label{eq_fvar}
\end{equation}
where $\langle F\rangle$ is the mean flux, $\sigma^2$ is the variance, and $\Delta^2$ is the mean square error. 
The uncertainty in $F_{\rm var}$ is expressed as (\citealt{Edelson2002}) 
\begin{equation}  
\sigma_{_{F_{\rm var}}} = \frac {1} {F_{\rm var}} \left(\frac {1}{2 N}\right)^{1/2} \frac {\sigma^2}{\langle F\rangle^2}, 
\end{equation}  
where $N$ represents the total number of observations. 
The resulting variability amplitudes are presented in Figure~\ref{fig_map2023} and listed in Table~\ref{tab_lcstat}, 
indicating that the broad He~{\sc ii} line has a larger fractional variation than the AGN continuum at 5100~\AA.

It should be noted that, in our previous work (\citealt{Lu2022}), we verified spectral calibration quality using NGC~5548 photometric light curves 
and estimated spectroscopic calibration precision using [O~{\sc iii}]~$\lambda5007$ fluxes. 
Since these examinations confirmed that the spectra are well calibrated by the above observing and calibration method, 
they were not repeated in this study. 

\begin{deluxetable*}{cccccccccccc}
  \tablecolumns{12}
  \tabletypesize{\scriptsize}
  \setlength{\tabcolsep}{10pt}
  \tablewidth{8pt}
  \tablecaption{Time Lags in Rest Frame\label{tab_lag}}
  \tablehead{
\colhead{}  &
\multicolumn{2}{c}{$F_{\rm He~II}~vs.~F_{\rm 5100}$}         &
\colhead{}  &
\multicolumn{2}{c}{$F_{\rm He~I}~vs.~F_{\rm 5100}$}          &
\colhead{}  &
\multicolumn{2}{c}{$F_{\rm H\gamma}~vs.~F_{\rm 5100}$}  &
\colhead{}  &
\multicolumn{2}{c}{$F_{\rm H\beta}~vs.~F_{\rm 5100}$}       \\ \cline{2-3} \cline{5-6} \cline{8-9} \cline{11-12}
\colhead{Season}                 &
\colhead{$\tau_{\rm He~II}$ (days)}                        &
\colhead{$r_{\rm max}$}       &
\colhead{}  &
\colhead{$\tau_{\rm He~I}$ (days)}                        &
\colhead{$r_{\rm max}$}       &
\colhead{}  &
\colhead{$\tau_{\rm H\gamma}$ (days)}                        &
\colhead{$r_{\rm max}$}       &
\colhead{}  &
\colhead{$\tau_{\rm H\beta}$ (days)}                        &
\colhead{$r_{\rm max}$}       
}
\startdata
2015 &$ 0.7_{-0.9}^{+1.1} $& 0.85 &&$ 4.4_{-1.4}^{+1.7} $& 0.80 &&$ 5.1_{-2.2}^{+2.3} $& 0.70 &&$  7.2_{-0.4}^{+1.3} $& 0.83 \\ 
2018 &$-0.3_{-2.1}^{+1.8} $& 0.92 &&$ 3.3_{-2.2}^{+1.5} $& 0.96 &&$ 4.3_{-4.9}^{+1.9} $& 0.86 &&$  7.0_{-3.4}^{+2.3} $& 0.94 \\
2019 &$ 0.8_{-1.1}^{+0.8} $& 0.94 &&$ 5.5_{-0.9}^{+2.5} $& 0.92 &&$ 4.5_{-1.3}^{+1.5} $& 0.90 &&$  8.9_{-1.1}^{+2.0} $& 0.94 \\
2020 &$ 1.1_{-1.0}^{+1.6} $& 0.63 &&  --                            & 0.26  &&   --                           & 0.11 &&$ 10.0_{-3.3}^{+3.3} $& 0.56 \\
2021 &$ 1.3_{-0.9}^{+1.1} $& 0.95 &&$ 5.5_{-3.3}^{+1.8} $& 0.88 &&$ 4.6_{-2.3}^{+1.2} $& 0.88 &&$  9.0_{-2.5}^{+1.9} $& 0.81 \\
2023 &$ 1.3_{-0.6}^{+1.6} $& 0.92 &&$ 2.3_{-2.1}^{+1.5} $& 0.74 &&$ 10.0_{-1.8}^{+2.0} $& 0.83 &&$  15.6_{-2.9}^{+2.6} $& 0.71
\enddata
\tablecomments{\footnotesize
Time lags between the variations of broad-line fluxes and AGN continuum strength at 5100~\AA\ are calculated. 
$r_{\rm max}$ represents the maximum correlation coefficient. 
The measurement results for the observing seasons of 2015, 2018, 2019, 2020 compiled from \cite{Lu2022} are included. 
}
\end{deluxetable*}

\begin{deluxetable*}{lccccccccc}
\tabletypesize{\scriptsize}
\setlength{\tabcolsep}{4pt}
\tablewidth{0pt}
\tablecaption{Broad H$\beta$ Line Widths and Supermassive Black Hole Masses \label{tab_mbh}}
\tablehead{
\multicolumn{4}{c}{Mean spectrum}         &
\colhead{}  &
\multicolumn{4}{c}{RMS spectrum}          &
\\ \cline{2-5} \cline{7-10} \\
\colhead{Season} &
\colhead{\shortstack{FWHM \\ (km s$^{-1}$) \\ (1)}} &
\colhead{\shortstack{$M_\bullet$ ($f=1.3\pm0.4$) \\  ($\times10^{8}M_{\odot}$) \\ (2)}} &
\colhead{\shortstack{$\sigma$ \\ (km s$^{-1}$) \\ (3)}} &
\colhead{\shortstack{$M_\bullet$ ($f=5.6\pm1.3$) \\  ($\times10^{8}M_{\odot}$) \\ (4)}} &
\colhead{}  &
\colhead{\shortstack{FWHM \\ (km s$^{-1}$) \\ (5)}} &
\colhead{\shortstack{$M_\bullet$ ($f=1.5\pm0.4$) \\  ($\times10^{8}M_{\odot}$) \\ (6)}} &
\colhead{\shortstack{$\sigma$ \\ (km s$^{-1}$) \\ (7)}} &
\colhead{\shortstack{$M_\bullet$ ($f=6.3\pm1.3$)\\  ($\times10^{8}M_{\odot}$) \\ (8)}}
}
\startdata
2015 & 11907$\pm$119 & $2.59^{+0.93}_{-0.81}$ & 4615$\pm$28 & $1.68^{+0.50}_{-0.40}$ && 9852$\pm$770 & $2.05^{+0.74}_{-0.64}$ & 5405$\pm$181 & $2.59^{+0.80}_{-0.65}$ \\
2018 & 11120$\pm$133 & $2.20^{+1.00}_{-1.25}$ & 3892$\pm$12 & $1.16^{+0.47}_{-0.62}$ && 10090$\pm$747 & $2.09^{+0.94}_{-1.19}$ & 4805$\pm$110 & $1.99^{+0.82}_{-1.07}$ \\
2019 & 10437$\pm$55  & $2.46^{+0.94}_{-0.81}$ & 3801$\pm$17 & $1.40^{+0.46}_{-0.37}$ && 10763$\pm$501 & $3.01^{+1.10}_{-0.92}$ & 4870$\pm$158 & $2.59^{+0.87}_{-0.71}$ \\
2020 & 9179$\pm$50   & $2.14^{+0.96}_{-0.96}$ & 3878$\pm$19 & $1.65^{+0.66}_{-0.66}$ && 10072$\pm$231 & $2.98^{+1.86}_{-1.86}$ & 5923$\pm$332 & $4.33^{+1.82}_{-1.81}$ \\
2021 & 8895$\pm$75   & $1.81^{+0.68}_{-0.75}$ & 3664$\pm$18 & $1.32^{+0.42}_{-0.48}$ && 8479$\pm$561  & $1.90^{+0.69}_{-0.77}$ & 4571$\pm$173 & $2.32^{+0.76}_{-0.86}$ \\
2023 & 9654$\pm$46   & $3.68^{+1.30}_{-0.13}$ & 4440$\pm$10 & $3.35^{+0.97}_{-1.00}$ && 10518$\pm$851 & $5.04^{+1.79}_{-1.84}$ & 5447$\pm$170 & $5.68^{+1.70}_{-1.76}$ \\
\hline
mean &-&$2.48\pm0.59$&-&$1.76\pm0.73$& &-&$2.85\pm1.09$&-&$3.25\pm1.32$ 
\enddata
\tablecomments{\footnotesize
The broadening-corrected line widths of broad H$\beta$, including FWHM and $\sigma_{\rm line}$, 
were measured from both the mean and rms spectra during the 2023 observing season, 
as well as updated measurements from the observing seasons of 2015, 2018, 2019, 2020, and 2021 (see \citealt{Lu2022} and Section~\ref{sec:lag}). 
Contamination from other blended components was removed through spectral fitting and decomposition. 
Supermassive black hole masses in Columns (2), (4), (6), and (8) were estimated using four types 
of H$\beta$ line widths (corresponding to Columns (1), (3), (5), and (7)) and their corresponding time lags 
(see Table~\ref{tab_lag}). 
The averaged supermassive black hole mass for each case is presented in the bottom row. 
The final average virial mass is $M_\bullet/10^{8}M_{\odot} = 2.58$, with a standard deviation of 1.12. 
}
\end{deluxetable*}

\subsection{Time Lag and Line width} \label{sec:lag}
The time lags of the broad He~{\sc ii}, He~{\sc i}, H$\gamma$, and H$\beta$ lines relative to the AGN continuum at 5100~\AA\ are 
calculated using the interpolation cross-correlation function (ICCF; \citealt{Gaskell1986,Gaskell1987,Sun2018}). 
Following \cite{Peterson2004}, the centroid of the ICCF above $0.8~r_{\rm max}$ is adopted as the time lag, 
where $r_{\rm max}$ is the maximum cross-correlation coefficient. 
Monte Carlo simulations with random subset sampling and flux randomization are used to construct the cross-correlation centroid distribution (CCCD), 
and the 15.87\% and 84.13\% quantiles of CCCD are used to estimate the uncertainties of the time lag. 

\begin{figure}[htp]
\includegraphics[width=0.49\textwidth]{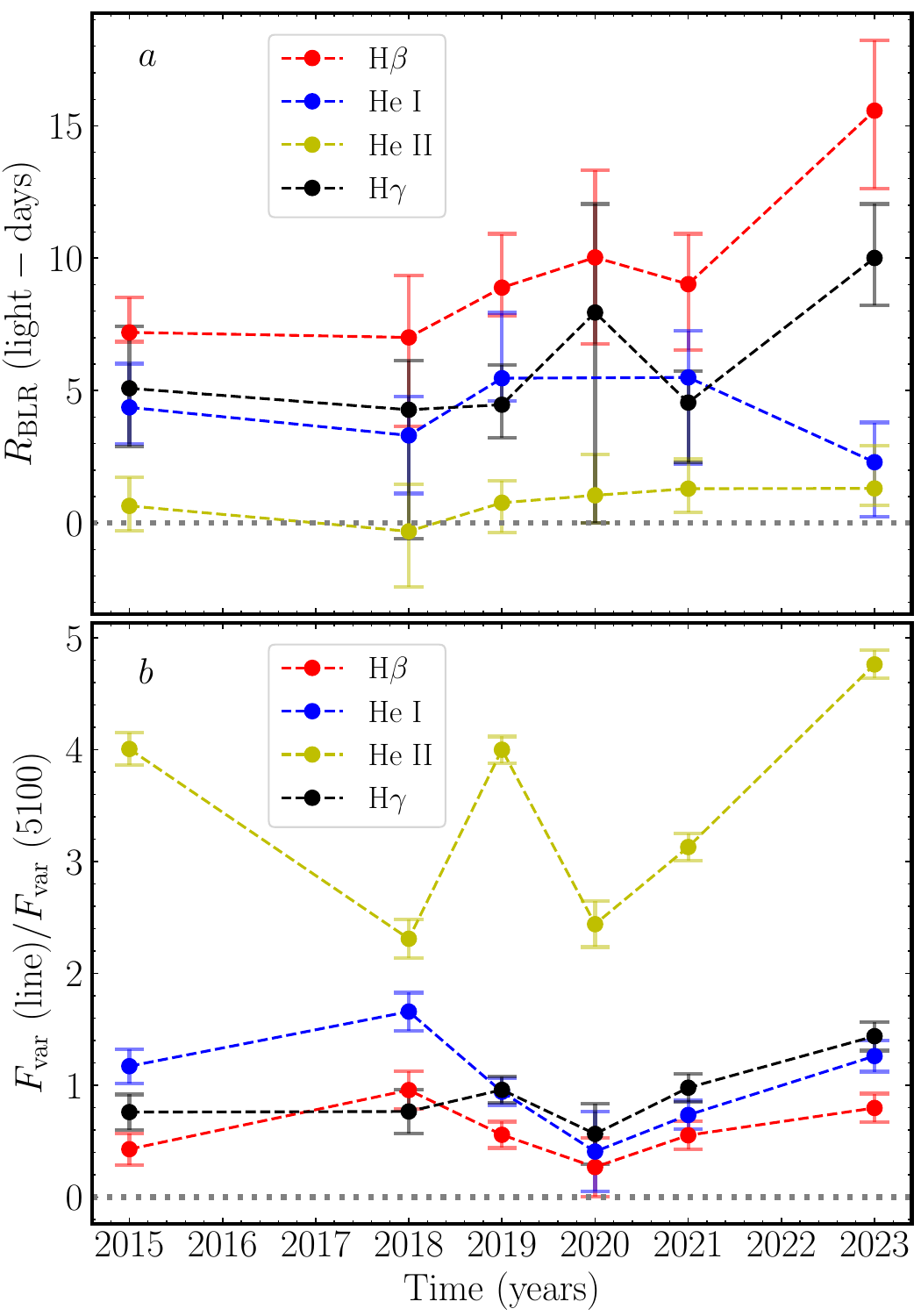}
\caption{
Radial stratification and line responsivity of the BLR in NGC~5548. 
Panel~({\it a}) presents the comparison of radii for broad-line emitters, 
which exhibits the properties of radial stratification. 
Panel~({\it b}) illustrates the comparison of responsivity for Balmer and Helium lines. 
}
\label{fig:blr}
\end{figure}

Cross-correlation analysis results are shown in Figure~\ref{fig_map2023}:  
(1) The autocorrelation function (ACF) of the AGN continuum light curve is displayed in panel~({\it aa}).  
(2) Cross-correlation functions (CCFs) between the AGN continuum and broad emission lines (red) along with their CCCDs (black) are shown in panels~({\it ba}$-${\it ea}).  
(3) Rest-frame time lags ($\tau_{\rm He~II}$, $\tau_{\rm He~I}$, $\tau_{\rm H\gamma}$, $\tau_{\rm H\beta}$) 
are marked by vertical dashed lines in CCF panels and listed in Table~\ref{tab_lag}. 
(4) Maximum cross-correlation coefficients are noted in CCF panels and also listed in Table~\ref{tab_lag}.  
Compared with the measurements from the past ten years (see Table 4 of \citealt{Lu2022} and this study), 
we find that the radius of the broad Balmer emission-line region in 2023 has increased by a factor of approximately two. 
A comparison with earlier measurements will be made in Section~\ref{sec:blrsize}. 
In contrast, the radius of the helium broad-line emission region shows no significant change 
during the observation period of Lijiang 2.4-meter telescope; for example, 
the radius of the broad He~{\sc ii} line emitters remains at approximately one light-day (also see top panel of Figure~\ref{fig:blr}). 

The broad-line width is characterized by either the full width at half maximum (FWHM) or line dispersion ($\sigma_{\rm line}$). 
In RM campaigns, these parameters are typically derived from both the mean and rms spectra. 
Notably, the broad Balmer lines of NGC~5548 display a double-peaked profile consistently across all spectroscopic monitoring periods. 
When the double peaks are well-resolved, the FWHM is determined as the wavelength separation between the blue-side and red-side peaks, 
following the method outlined in \cite{Peterson2004}. 

For this observing season, we conducted Monte Carlo simulations using random subset selection to generate 200 mean and rms spectra 
for the broad H$\beta$ line. From these spectra, we derived four line-width distributions: 
FWHM (mean), $\sigma_{\rm line}$ (mean), FWHM (rms), and $\sigma_{\rm line}$ (rms). 
The average values and standard deviations of each distribution were adopted as the best estimates and uncertainties for the line widths. 
Using this approach, we refined our previous measurements of the broad H$\beta$ line widths (\citealt{Lu2022}). 

In practice, the observed broadening of emission lines results from a combination of instrumental and atmospheric (seeing) effects. 
To quantify these contributions, we compared the FWHM of [O~{\sc iii}]~$\lambda$5007 measured from a high-resolution spectrum 
(410~km~s$^{-1}$; \citealt{Whittle1992}) with the averaged FWHM (810~km~s$^{-1}$) obtained from our spectroscopic campaign. 
Assuming a Gaussian line profile for [O~{\sc iii}], 
this analysis yields an instrumental broadening of 699~km~s$^{-1}$ (FWHM) or 297~km~s$^{-1}$ ($\sigma_{\rm line}$).  

The corrected line widths, accounting for both instrumental and atmospheric broadening, are summarized in Table~\ref{tab_mbh}. 
This table includes updated measurements for the broad H$\beta$ line 
from the observing seasons of 2015, 2018, 2019, 2020, and 2021 (\citealt{Lu2022}). 
These refined values will serve as critical inputs for estimating the mass of the supermassive black hole. 

\section{Supermassive Black Hole Mass}\label{sec:mass}
Using RM measurements of the broad H$\beta$ line's width and its time lag, 
we can estimate the SMBH mass via the virial equation:  
\begin{equation}
M_\bullet = f\frac{c\tau_{\rm H\beta} V^{2}}{G}, 
\label{eqn:mass}
\end{equation}
where $c$ is the speed of light, $\tau_{\rm H\beta}$ is the H$\beta$ time lag, $c\tau_{\rm H\beta}$ means the BLR radius (or size), 
$G$ is the gravitational constant, 
$V$ is the line width (either FWHM or $\sigma_{\rm line}$), representing the BLR velocity in line of sight, 
and $f$ is a dimensionless virial factor accounting for BLR geometry, kinematics, and inclination. 
\citet{Ho2014} showed that $f$ depends on the host galaxy's bulge type (classical bulge or pseudobulge). 
Since NGC~5548 hosts a classical bulge (\citealt{Ho2014}), the classical bulge-dependent $f$ are adopted in the next calculation (see Table~\ref{tab_mbh}). 

As stated in Section~\ref{sec:lag}, the H$\beta$ line widths (either FWHM or $\sigma_{\rm line}$) can be measured from the mean and rms spectra, 
therefore we estimated the virial SMBH mass in four cases using the H$\beta$ line widths and time lags, as presented in Table~\ref{tab_mbh}. 
For each case, the average SMBH mass was calculated and listed in the bottom row of Table~\ref{tab_mbh}. 
Then using the virial masses of four cases, we compute the final average virial mass to be $(2.58 \pm 1.12) \times 10^8 M_{\odot}$. 
The stellar velocity dispersion of the classical bulge in NGC~5548 is $\sigma_* = 195 \pm 13~{\rm km~s^{-1}}$ (\citealt{Woo2010}), 
based on the $M_{\rm BH}-\sigma_*$ relation from \cite{Kormendy2013}, 
we obtained $M_{\rm BH}|_{\sigma_*} = (2.75 \pm 0.88) \times 10^8 M_{\odot}$ (\citealt{Lu2016}). 
These results show that virial mass of SMBH can be reliably determined across all four line width measurement methods considered, 
and confirm the remarkable consistency between the RM-based virial mass and the estimate derived from the $M_{\rm BH}-\sigma_*$ relation. 

\begin{figure*}[htp]
\center
\includegraphics[width=0.49\textwidth]{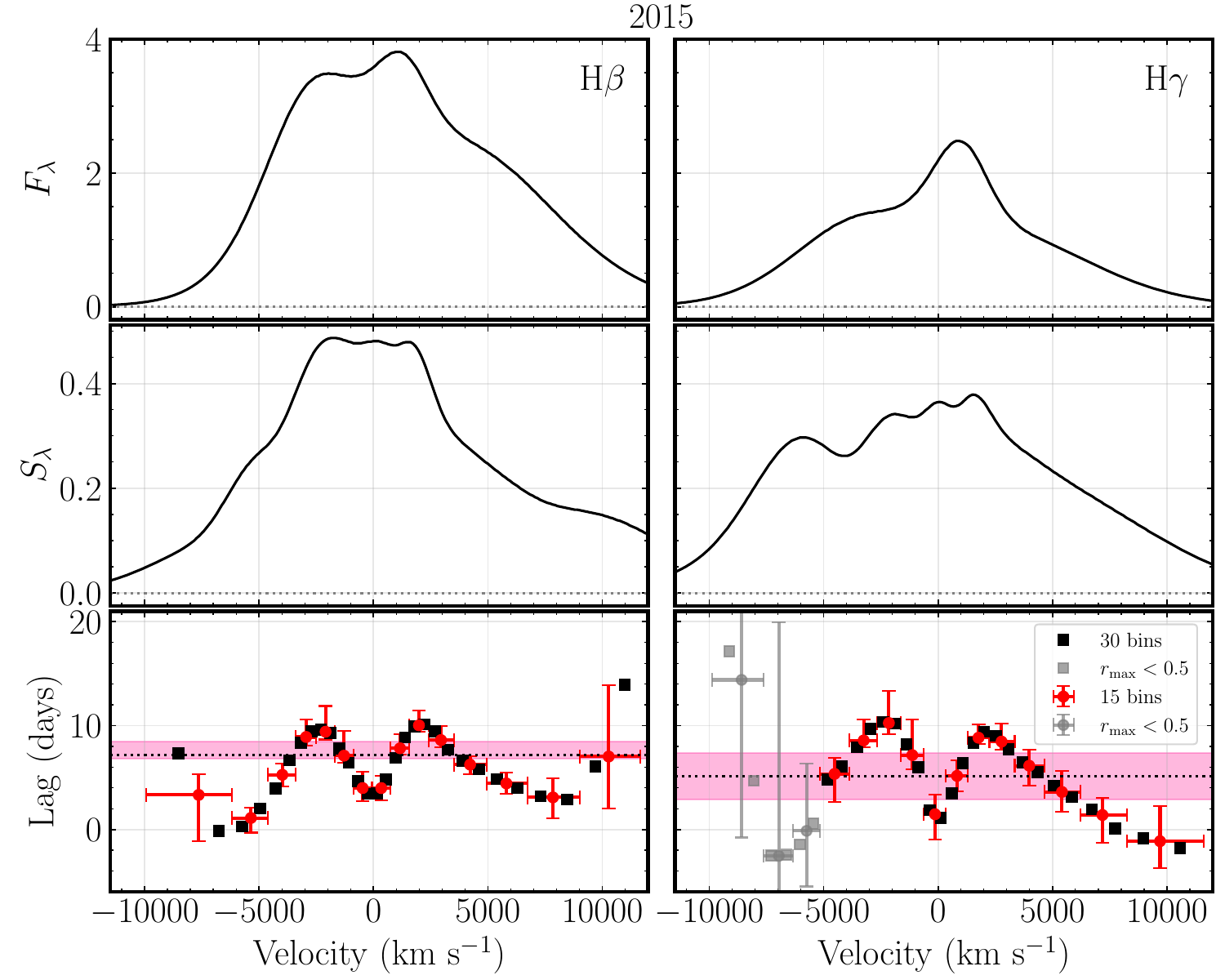}
\includegraphics[width=0.49\textwidth]{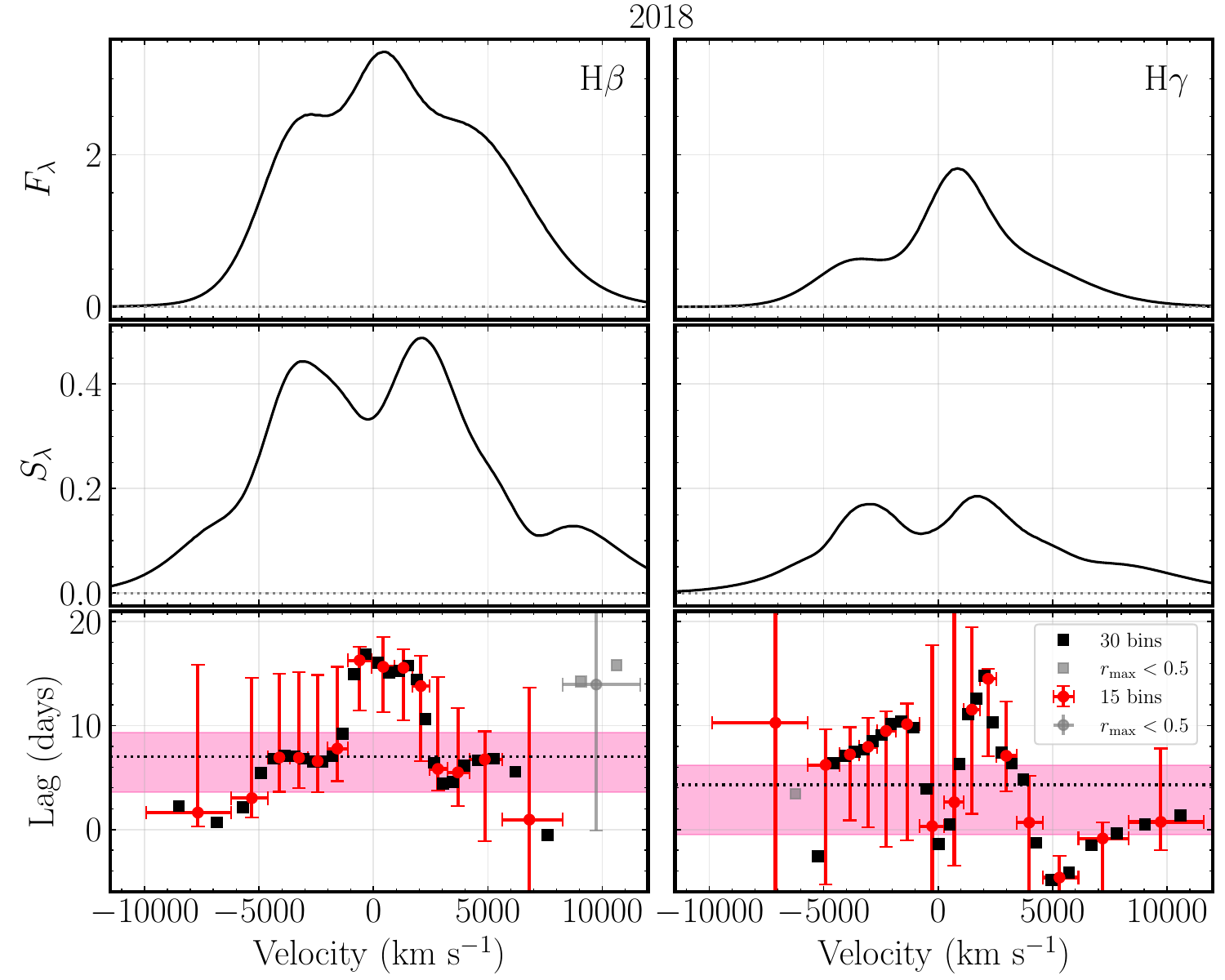}\\
\includegraphics[width=0.49\textwidth]{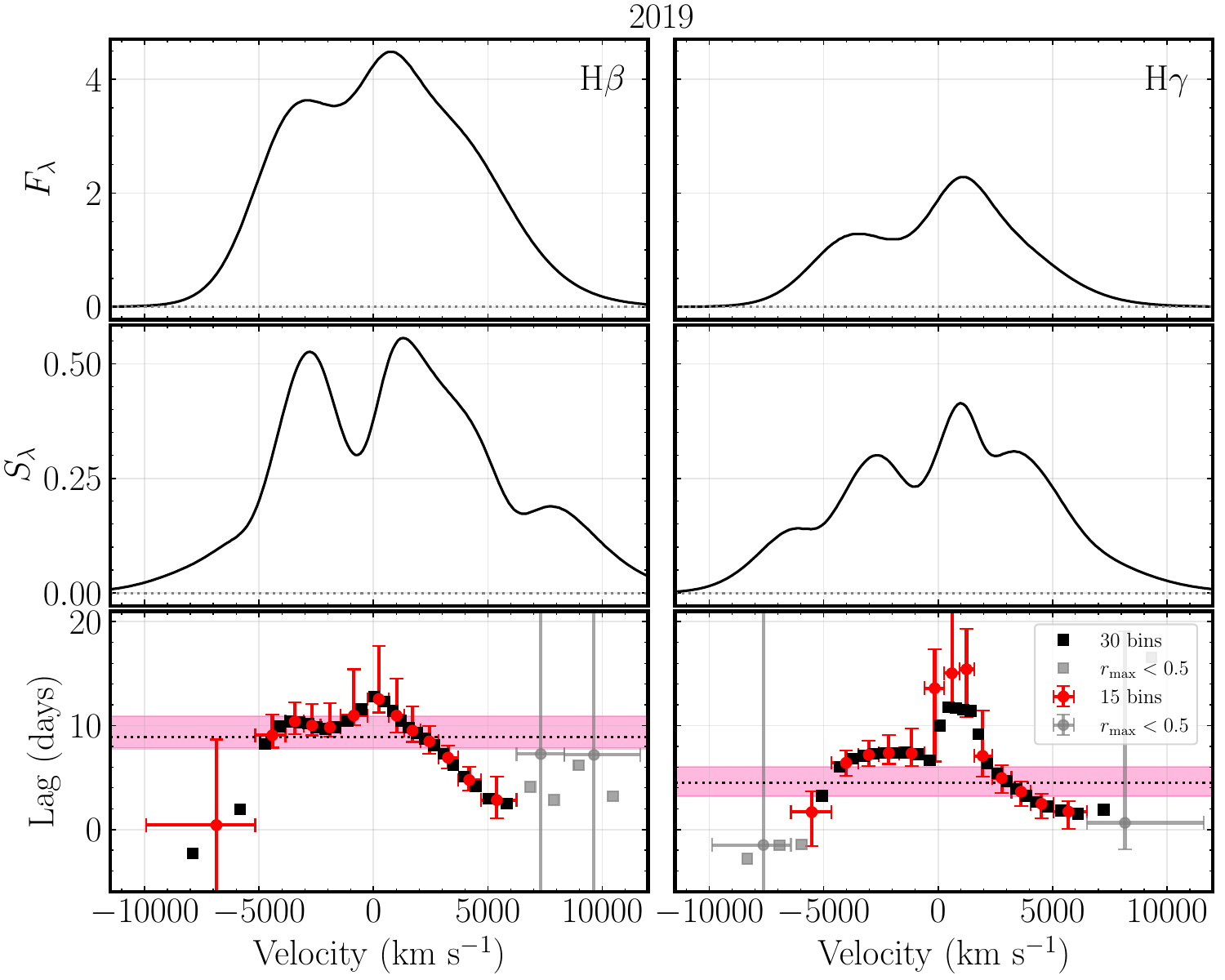}
\includegraphics[width=0.49\textwidth]{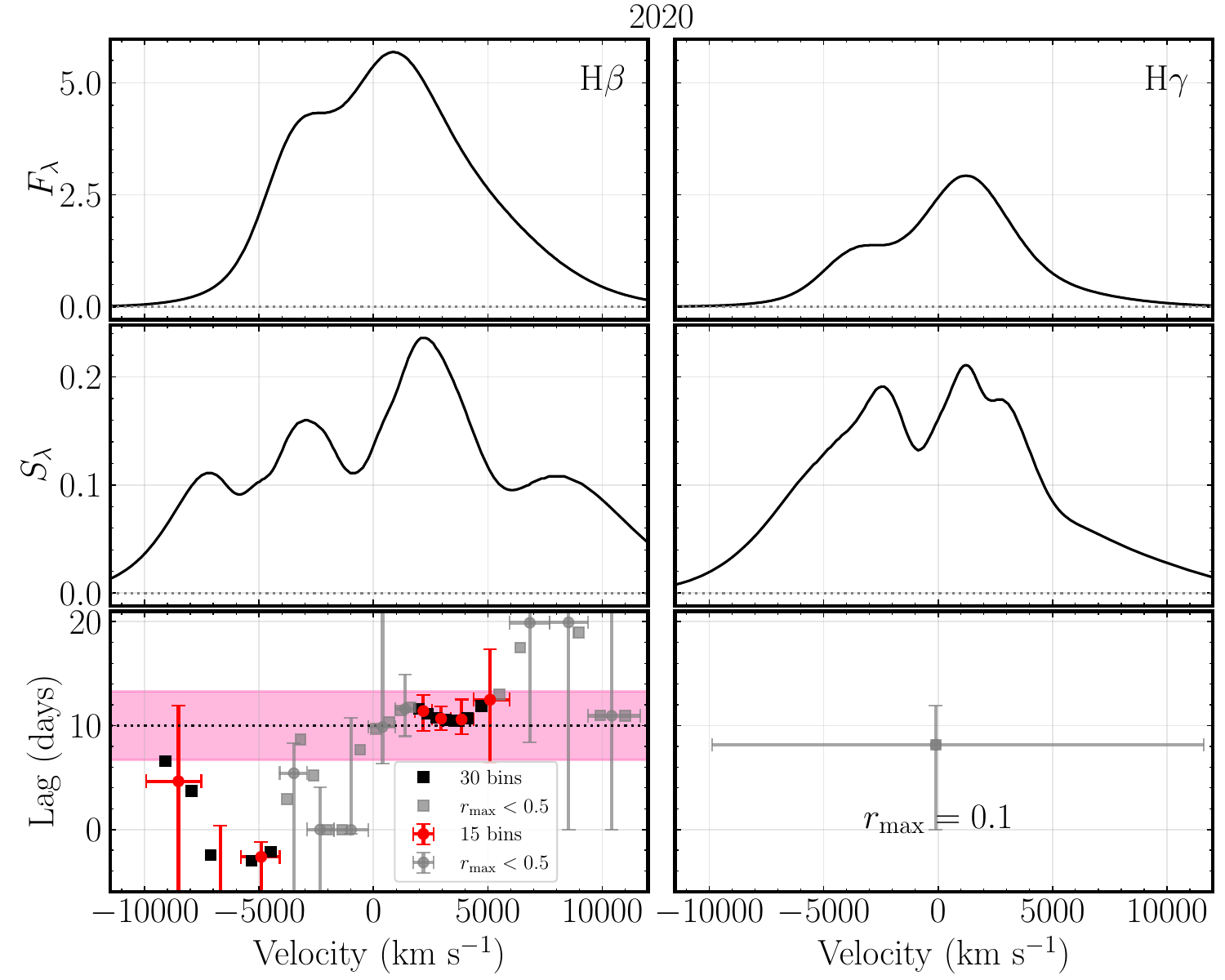}\\
\includegraphics[width=0.49\textwidth]{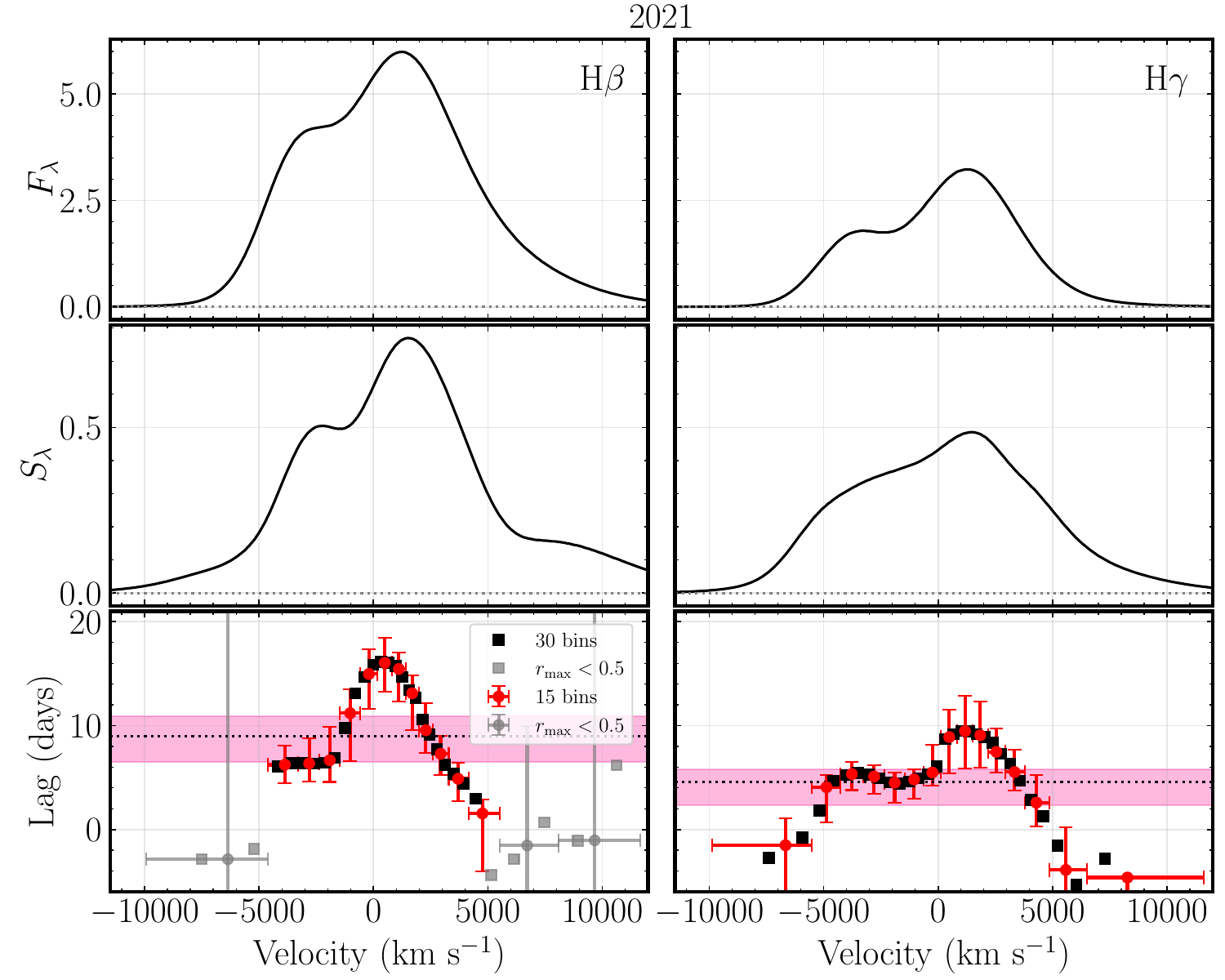}
\includegraphics[width=0.49\textwidth]{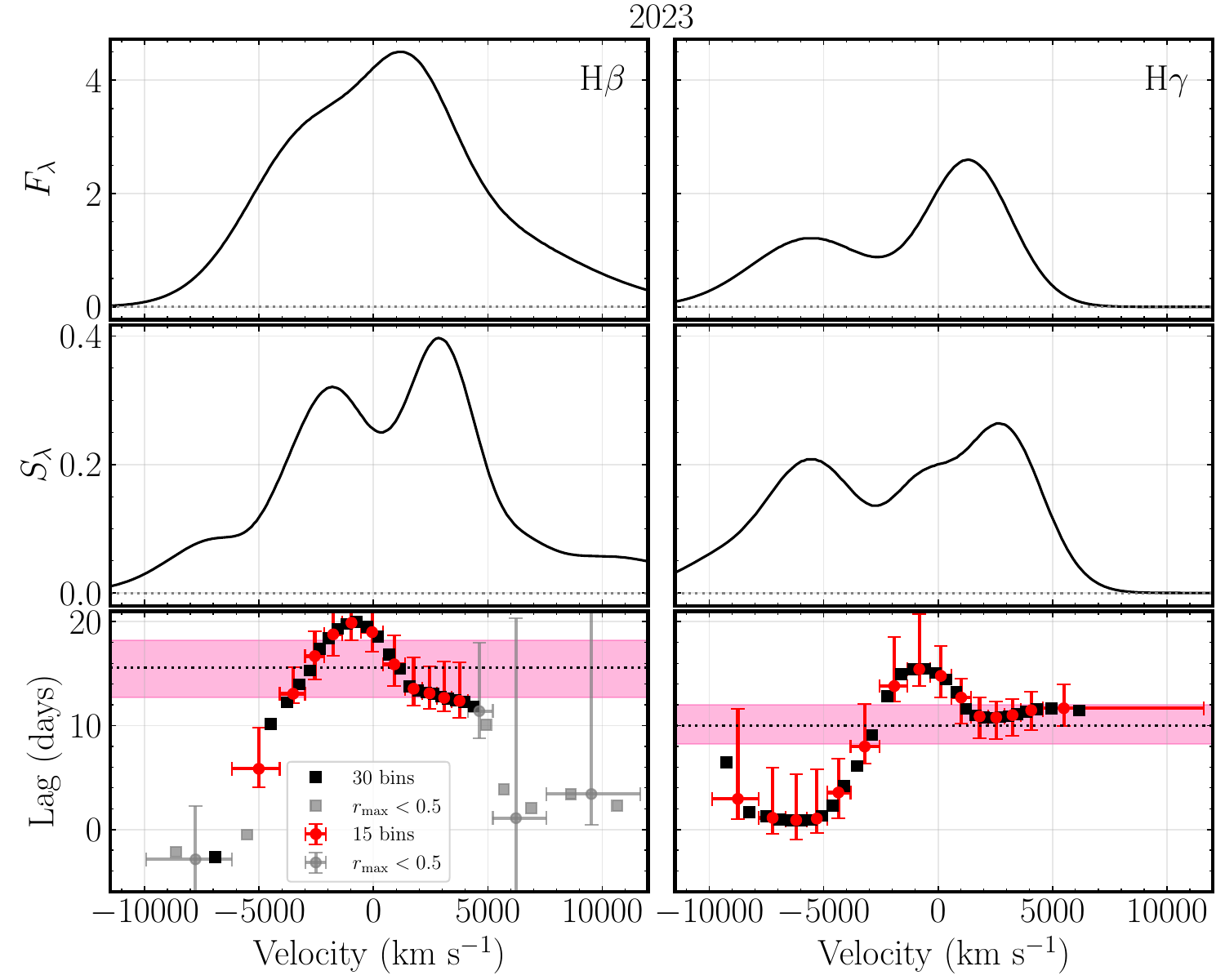}
\caption{
Velocity-resolved RM results for the 2023 season (bottom right). 
For comparison, updated results from previous seasons (2015, 2018, 2019, 2020, 2021) are shown. 
The top panel shows the mean spectrum; the middle panel shows the rms spectra of broad H$\gamma$ (right) and H$\beta$ (left). 
The bottom panel displays centroid time lags versus line-of-sight velocity, based on 15 (circles) and 30 (squares) equal-flux velocity bins. 
Horizontal error bars represent half the bin width. Gray symbols indicate bins with maximum correlation coefficients below 0.5. 
No RM detection was obtained in 2020 for H$\gamma$ due to weak variability. 
The horizontal dotted line and pink band mark the average lag and its uncertainty for each season.  
}
\label{fig:ljmaps}
\end{figure*}

\section{Broad-line Region Properties of NGC~5548}\label{sec:blr}
To date, six RM observations of NGC~5548 have been conducted 
using the 2.4-meter telescope (see \citealt{Lu2022} and this work), 
spanning the observing seasons of 2015, 2018, 2019, 2020, 2021, and 2023. 
In this section, we combine the new measurement presented in this work with 
the previously reported RM results to investigate the properties of the BLR. 

\subsection{Radial Ionization Stratification and Line Responsivity}
It is well-established that radial ionization stratification occurs due to ionization-energy effects (\citealt{Collin1988}). 
Ions with higher ionization potentials, such as 54.4 $\mathrm{eV}$ for He~{\sc ii}~$\lambda4686$, 
require a stronger ionizing radiation field and are thus produced more efficiently closer 
to the central engine than ions with lower ionization potentials, such as 13.6 $\mathrm{eV}$ for H$\beta$. 
Our results confirm this fundamental principle: in all six of RM measurements for NGC~5548, 
the He~{\sc ii} time lags are consistently shorter than those of the Balmer lines (H$\beta$ and H$\gamma$, see Figure~\ref{fig:blr}), 
aligning with theoretical predictions and previous findings (e.g., \citealt{Bentz2010, Hu2020}). 

A more nuanced picture of the BLR, particularly for the Balmer lines, 
emerges from the Locally Optimally Emitting Clouds (LOC) model (\citealt{Baldwin1995}). 
This model posits that the BLR consists of a vast, 
chaotic assemblage of clouds with a wide range of densities and distances from the central engine. 
The observed spectrum is not produced by clouds with fine-tuned parameters, 
but is the integrated result of selection effects: for each emission line, 
there is a corresponding region in the density$-$ionizing flux parameter space where its reprocessing efficiency is maximized. 

LOC models assume that the ionizing photon flux, $\Phi (H)$, is directly related to the distance from the central source, $r$, 
by $\Phi (H)\propto L/r^{2}$, where $L$ is the luminosity. 
Therefore, the optimal emitting conditions for a given line correspond to a characteristic radius with in the BLR. 
This model naturally predicts a radial stratification of the BLR, where different emission lines originate, on average, from different distances, 
based on their atomic physics (ionization potential, thermalization density). 
Our finding of H$\beta$ time lags being consistently longer than those of H$\gamma$ (see Figure~\ref{fig:blr}), 
to some extent, confirm that the characteristic radius for optimal H$\beta$ emission is larger than that for H$\gamma$ 
(e.g., \citealt{Bentz2010,Feng2021}). 
This stratification is a natural outcome of the different atomic physics governing each line's emissivity. 
The role of optical depth is secondary, it can modulate the line response function through effects like anisotropic emission 
but does not cause the primary lag difference (e.g., \citealt{Rees1989,Korista2004}). 
From the six of RM measurements for the time lags in 2015 to 2023, 
we found that H$\beta$ time lags are consistently longer than those of H$\gamma$ (see Figure~\ref{fig:blr}). 
These results demonstrate that radial ionization stratification of the BLR gas is persistent and stable. 

The efficiency with which BLR gas converts a varying ionizing flux into line flux corresponds to the responsivity ($\eta$) 
of an emission line (see \citealt{Korista2004}), which is often assessed by measuring the ratio of the fractional variation 
$F_{\rm var}(\rm line)$ for the emission line relative to $F_{\rm var}(\rm 5100)$ of the AGN continuum (Figure~\ref{fig:blr}). 
Our results reveal a clear pattern of $\eta_{\rm He~II} \gg \eta_{\rm He~I} \approxeq \eta_{\rm H\gamma} > \eta_{\rm H\beta}$. 
This sequence is fully consistent with photoionization calculations (\citealt{Korista2004}). 
Within the LOC framework, 
this pattern is interpreted that higher responsivity gas (like He~{\sc ii} emitters) is located at the optimal emission region 
(typically closer to central ionizing source), 
where a small change in the ionizing continuum leads to large variability of the emission-line flux. 

\begin{figure*}[ht!]
\center
\includegraphics[width=0.97\textwidth]{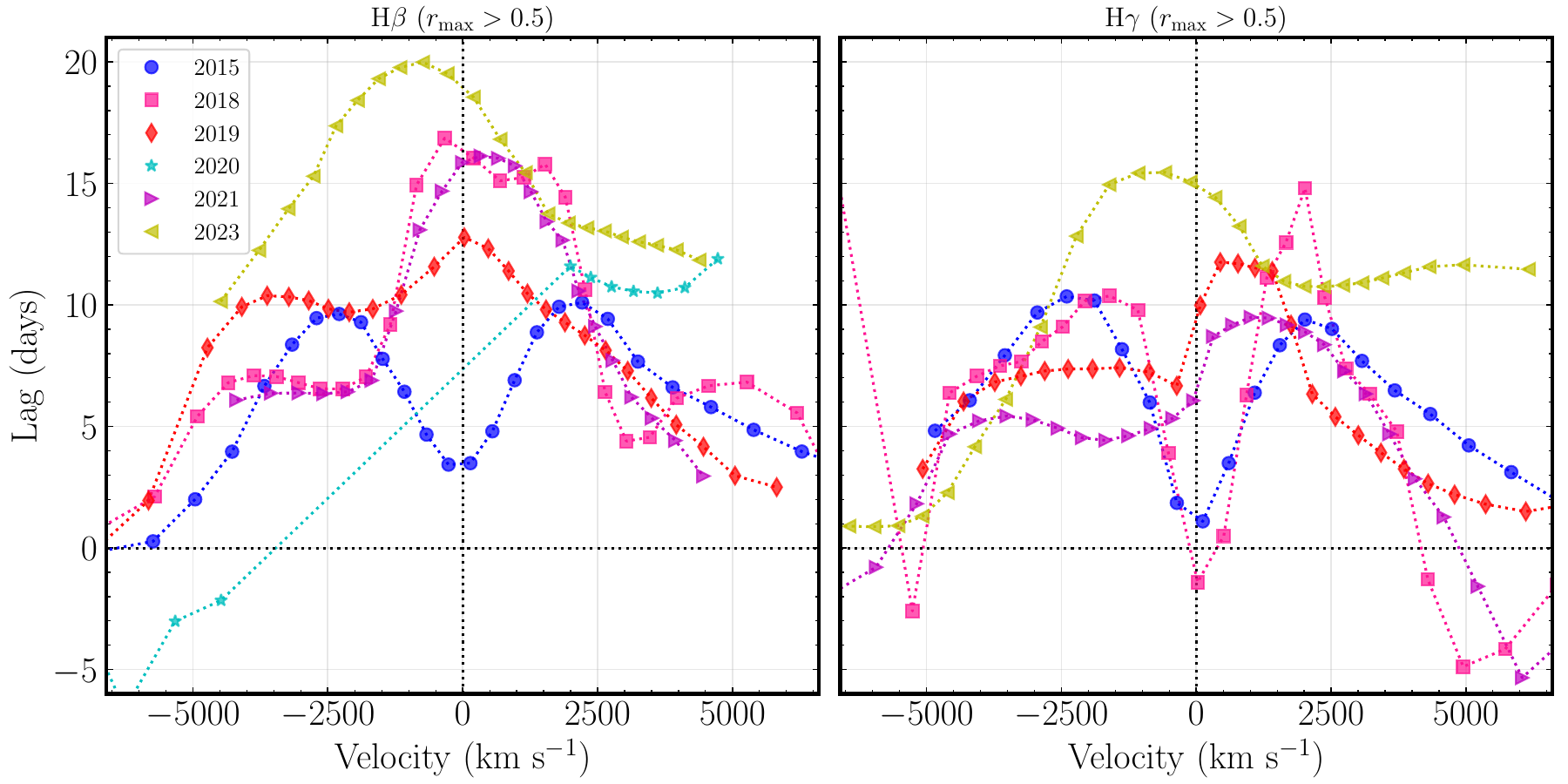}
\caption{
The temporal variations of H$\beta$ and H$\gamma$ VRLPs are systematically compared 
across six observing seasons (2015, 2018, 2019, 2020, 2021, and 2023) using 30 equal-flux velocity bins, 
with lags below a maximum correlation coefficient of 0.5 excluded. 
Vertical and horizontal dotted lines indicate the line core at zero velocity and zero time lag, respectively. 
}
\label{fig:mapscomp}
\end{figure*}

\subsection{Kinematics and its Temporal Variations}\label{sec:vr}
We performed velocity-resolved RM analysis using the net broad H$\gamma$ and H$\beta$ lines obtained in Section~\ref{sec:lc}. 
Following previous studies (e.g., \citealt{Denney2010,Bentz2009,Grier2013,Du2016,Lu2016,Pei2017,Lu2022}), 
we first calculated the mean and rms spectrum of the net broad emission lines (i.e., continuum-subtracted). 
We then defined a broad emission-line window spanning 4700$-$5050~\AA\ for H$\beta$, 
and 4197$-$4508~\AA\ for H$\gamma$ to encompass the broad wings in both the mean and rms profiles. 
Within each window, we divided the broad line into velocity bins of equal flux, 
constructed velocity-resolved light curves by integrating the fluxes in each bin, 
and measured the time lags between these light curves and the AGN continuum at 5100~\AA~using 
the procedures outlined in Section~\ref{sec:lag}. 
Except for the observing season of 2023, the results from previous seasons (2015, 2018, 2019, 2020, 2021; see \citealt{Lu2022}) 
are updated using the same method in the same measurement window, 
as demonstrated in Figure~\ref{fig:ljmaps}. 
The top and middle panels display the mean and rms spectra of broad H$\gamma$ (right) and H$\beta$ (left) lines, respectively. 
The bottom panels show the velocity-resolved lag profiles (VRLPs), 
where the time lags are plotted as a function of line-of-sight velocity (hereafter H$\beta$ VRLP and H$\gamma$ VRLP). 
Such a combination chart corresponds to one observing season. 
We confirmed that the VRLPs derived from the fitted broad lines are consistent with those measured from 
the broad lines after subtracting blended components from the calibrated spectra. 
The broad helium lines were excluded from this analysis due to their very weak fluxes. 

A comparative analysis of the VRLPs across different seasons reveals a temporal variation in the BLR kinematics of NGC 5548, 
especially in the observing seasons of 2015, 2018, 2019, 2021, 2023 (see Figure~\ref{fig:mapscomp} for a comparison). 
In 2007, broad Balmer line shows symmetric VRLP, with the shortest lags in the line wings, 
consistent with a Keplerian motion of the BLR under isotropic illumination (\citealt{Bentz2009,Welsh1991}). 
In 2008, broad Balmer line shows no clear structure in VRLP (\citealt{Bentz2009}). 
In 2012 (\citealt{DeRosa2018}), 2014 (\citealt{Pei2017}), and 2015 (\citealt{Lu2016,Lu2022} and here), 
both H$\beta$ (and H$\gamma$) exhibit a symmetric `M'-shaped structure in VRLP, 
with the shortest lags in the high-velocity wings, where high-velocity gas resides closer to the SMBH. 
According to the theoretical model of the BLR kinematics \citep{Welsh1991,Goad1996} 
and geometric suggestions \citep{McLure2002,Bisogni2017,Risaliti2011}, 
the `M'-shaped structure of VRLP arises from a virialized, disk-like BLR under anisotropic illumination. 
By 2018, the H$\gamma$ VRLP shows a distorted `M'-shaped structure, 
while H$\beta$ VRLP becomes less structured. 
In 2019 and 2021, the VRLPs display a general the red-leads-blue trend, 
though the lags in the red and blue wings are comparable (also see \citealt{Lu2022}), 
possibly indicating a combination of infalling and virialized motions 
(e.g., \cite{Gaskell1988,Horne2004,Xiao2018,Gaskell2013}) or a switch between these motions. 
While, the H$\beta$ VRLP in 2020 appears to show a blue-leads-red trend, consistent with outflow of the BLR gas.  
Notably, in 2023, both H$\beta$ and H$\gamma$ VRLPs show a general blue-leads-red signature, 
with longer lags in the red wing and the maximum lag shifted to the blue side of the line core. 
This pattern could stem from a mixture of outflow and 
virialized motions, consistent with theoretical models (\citealt{Grier2013,Welsh1991}). 
Alternatively, variable obscuration by accretion disk winds may lead to anisotropic illumination, generating complex VRLP signatures 
(e.g, \citealt{Mangham2017,Mangham2019,Gaskell2018,Dehghanian2019}). 

These results collectively demonstrate that the BLR kinematics in NGC 5548 are dynamic and evolve over time, 
corroborating previous findings of structural and kinematic changes in this object (\citealt{Xiao2018, Liss2022}). 
The observed variations, from virialized motions to signatures of inflow/outflow, highlight the complex nature of the BLR 
and its response to changes in accretion activity. A detailed dynamical modeling analysis of these data, 
using the methods of \citealt{Pancoast2011} and \citealt{Li2013} (also see \citealt{Pancoast2014,Li2018,Williams2020}), 
will be presented in Xi et al. (in preparation). 

\begin{figure*}[ht!]
\center
\includegraphics[width=0.97\textwidth]{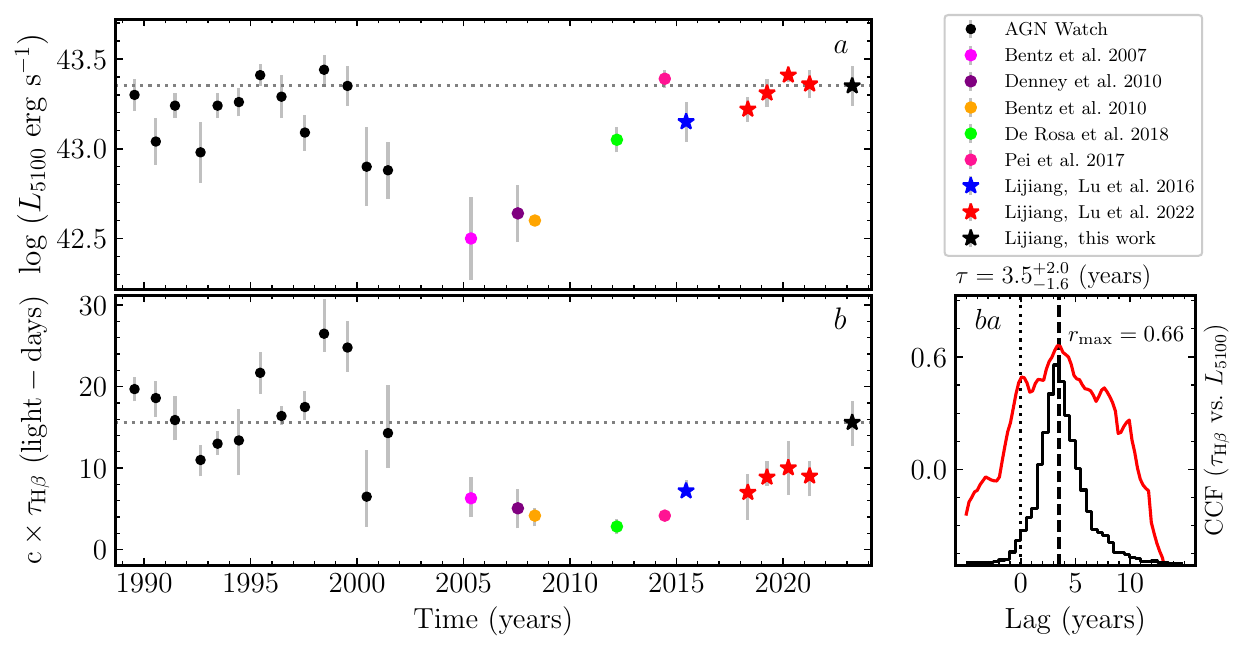}
\caption{
Variations of the optical luminosity ({\it a}) and the BLR radius ({\it b}) for NGC~5548 measured between 1988 and 2023, 
and the result of cross-correlation analysis (panel {\it ba}). 
The different symbols with different colors are used in the left panels to distinguish the different campaigns, 
The horizontal dotted lines in panels (a) and (b) mark the optical luminosity and BLR radius positions for the 2023 observing season, 
Both parameters in 2023 have already reached levels comparable to their historical maxima.  
In panel ({\it ba}), the red solid curve is the CCF between the BLR radius and the optical luminosity ($r_{\rm max}=0.66$), 
and the black histogram is the cross-correlation centroid distribution (CCCD), 
the vertical dashed line marks the measured time lag of $3.5^{+2.0}_{-1.6}$~years, 
the vertical dotted lines are reference lines of zero time lag. 
}
\label{fig:blrevo}
\end{figure*}

\subsection{BLR Radius Lags the Optical Luminosity} \label{sec:blrsize}
Using the mean continuum flux at 5100~\AA\ from the observing season of 2023 (see Table~\ref{tab_lcstat}), 
we compute the optical luminosity of $L_{\rm 5100}$, 
adopting the same cosmological parameters as those in \cite{Lu2022}. 
Then we examine the long-term variation of the optical luminosity 
and BLR radius ($R_{\rm BLR}=c\tau_{\rm H\beta}$) over a 35-year observational baseline, as shown in Figure~\ref{fig:blrevo}. 
The data show a sustained decrease in both parameters since 1998, lasting approximately 5 to 10 years, 
with minimum values reached around 2005 and 2010, respectively, 
followed by a gradual recovery over the subsequent decade to pre-decline levels. 
A review of historical observations reveals that NGC 5548 underwent 
a significant weakening of its optical continuum and broad Balmer line around 2005 (see \citealt{Bentz2009}), 
with the broad H$\beta$/[O~{\sc iii}] ratio approaching 0.33 (i.e., a Type 1.8 AGN; \citealt{Winkler1992}), 
a characteristic signature of a changing-look AGN. 
After this transition, the optical luminosity recovered to its historical maximum by $\sim$2020 
(see the horizontal dotted line in panel {\it a}), 
while the BLR radius, within measurement uncertainties, approached its typical historical value by 2023 
(see the horizontal dotted line in panel {\it b}). 

Building on our previous work of \cite{Lu2022}, 
we conduct a time-series analysis of both parameters using the methodology outlined in Section~\ref{sec:lag}. 
As shown in panel ({\it ba}) of Figure~\ref{fig:blrevo}, when the 2023 observations are incorporated, 
the measured time lag between $R_{\rm BLR}$ and $L_{\rm 5100}$ 
is consistent with the value reported by \cite{Lu2016}, at $3.5^{+2.0}_{-1.6}$~years. 
These results are consistent with the scenario where the central SMBH resumed its accretion activity, 
the BLR has largely reverted to its initial (predecline) configuration. 
In this case, the results suggest an underlying link between changes in accretion state and the structural response of the BLR. 

However, the 3.5-year lag between BLR radius and optical luminosity 
is comparable to the BLR dynamical timescale of $\sim$2~years (\citealt{Lu2016,Lu2022}). 
In practice, the observed BLR radius is actually an response-averaged value over the whole BLR, 
while radiation of central ionizing source induces changes in emissivity. 
Additionally, the BLR kinematics could be jointly controlled by radiation pressure and SMBH gravity. 
Collectively, the delayed response of the BLR radius to optical luminosity plausibly implicates 
the potential role of radiation pressure (also see \citealt{Lu2016}). 

On the other hand, this timescale is comparable to the potential rotation or precession period of inner disk structures ($\sim$2 years), 
as indicated by the ``barber pole" pattern revealed in MEMEcho residuals (\citealt{Horne2021}, also see Section~\ref{sec:intro}). 
To detect periodic signatures of accretion disk rotation or precession in the continuum variability and emission line profiles, 
we attempted a periodicity analysis using six-season reverberation mapping data from Lijiang, 
spanning nearly a decade of observations. 
The dataset includes measurements of the AGN continuum at 5100~\AA, the broad H$\beta$ line width, 
the peak separation and strength of the double-peaked H$\beta$ emission line. 
Nevertheless, no significant periodic signals are detected in the current dataset. 
Continued monitoring of NGC 5548 is therefore necessary to further investigate dynamical processes in its nuclear region. 

\section{Summary} \label{sec:sum}
NGC~5548 is the best-observed active galactic nucleus (AGN) with long-term reverberation mapping (RM) monitoring. 
It serves as an ideal case for studying the structure and evolution of the broad-line region (BLR). 
This paper presents the latest results from our ongoing long-term reverberation mapping campaign on the nearby Seyfert galaxy NGC 5548, integrating the 2023 observing season with our previous datasets. The main results are summarized as follows: 

\begin{enumerate}
\item
We successfully obtained 74 spectroscopic observations in 2023. 
Through spectral decomposition, we generated light curves for the continuum and broad emission lines. 
The cross-correlation analysis provided robust time lag measurements for 
broad He~{\sc ii}, He~{\sc i}, H$\gamma$, and H$\beta$ emission lines, 
confirming the expected sequence of $\tau_{\rm He~II} < \tau_{\rm He~I} < \tau_{\rm H\gamma} < \tau_{\rm H\beta}$. 

\item
Our measurements from six-season RM observation consistently show that the time lags of the broad emission lines follow 
the sequence $\tau_{\rm He~II} < \tau_{\rm He~I} \lesssim \tau_{\rm H\gamma} < \tau_{\rm H\beta}$. 
This firmly establishes the presence of a radially ionized stratified BLR. 
This result is consistent with the predictions of the Locally Optimally Emitting Clouds (LOC) model. 

\item
Using the H$\beta$ time lag and line width, 
we calculated a virial mass for the supermassive black hole of 
$M_\bullet/10^8M_\odot = 2.6$ with a standard deviation of $1.1$, 
which is in excellent agreement with the mass estimated from the $M_{\rm BH}-\sigma_*$ relation, 
reinforcing the reliability of RM-based mass measurements. 

\item
The velocity-resolved RM analysis reveals a significant temporal evolution in the BLR kinematics. 
The VRLPs transitioned from symmetric `M-shaped' profiles (e.g., 2015) indicative of a disk-like BLR, 
to `red-leads-blue' signatures (2019, 2021) suggesting infall, and finally to a `blue-leads-red' pattern in 2023, 
indicative of outflow motions. 
This demonstrates that the BLR is a dynamically evolving region, responsive to changes in the central engine. 

\item
Following an extreme decline in optical luminosity and a substantial contraction of the BLR radius, 
the optical luminosity recovered to its historical maximum by $\sim$2020, 
while the BLR radius returned to its typical historical value by $\sim$2023. 
Furthermore, the long-term (35-year) trend reveals that variations in the BLR radius lag 
behind changes in optical luminosity by approximately 3.5 years. 
This delayed response indicates a strong coupling between accretion activity and the structure of the broad-line region, 
likely driven by radiation pressure or associated with dynamical processes in the inner accretion flow, 
including possible disk precession. 
\end{enumerate}

In conclusion, our decade-long RM campaign on NGC 5548 demonstrates that 
its BLR is a robust yet dynamically evolving structure. The observed kinematic changes 
and the delayed response of the BLR to continuum variations provide crucial insights into 
the accretion physics in active galactic nuclei. Continued monitoring and detailed dynamical 
modeling of this key archetype will be essential for further unraveling the intricacies of the central regions of AGNs. 

\section*{Acknowledgments}
We gratefully acknowledge the referee for insightful report that enhanced the quality of the paper. 
This work is supported by the National Key R$\&$D Program of China with No. 2021YFA1600404, 
the National Natural Science Foundation of China (NSFC-12573020, 12073068). 
K.X.L. acknowledges financial support from 
the Youth Innovation Promotion Association of Chinese Academy of Sciences (2022058), 
the the Young Talent Project of Yunnan Province, the Yunnan Province Foundation, 
and the science research grants from the China Manned Space Project with No. CMS-CSST-2025-A07). 
L.X. acknowledges financial support from the Young Talent Project of Yunnan Province 
and the Yunnan Province Foundation (202401AT070138). 
We acknowledge the support of the staff of the Lijiang 2.4~m telescope. 
Funding for the telescope has been provided by Chinese Academy of Sciences and the People’s Government of Yunnan Province.

\end{document}